\documentclass{mem}
\usepackage{natbib}\usepackage{txfonts}\usepackage{balance}
\usepackage{flushend}
\usepackage{graphicx}
\usepackage[breaklinks,pdftex]{hyperref}
\idline{0}{1}
\begin{document}
\title{
The Firmamento platform}
\subtitle{A tool for blazar discovery and multimessenger research}
\author{
P.\ Giommi\inst{1,2} on behalf of the Firmamento development team}
\institute{
Center for Astrophysics and Space Science (CASS), New York University-Abu-Dhabi, P.O. Box 129188 Abu Dhabi, United Arab Emirates
\and
Associated to INAF, Brera Astronomical Observatory, via Brera, 28, I-20121 Milano, Italy
\email{paolo.giommi@nyu.edu}\\
}
\authorrunning{Giommi}
\titlerunning{Firmamento}
\date{Received:  (Day-Month-Year); Accepted:  (Day-Month-Year)}

\abstract{
This paper introduces Firmamento, a new online platform designed for astronomical research, particularly for studying blazars and other multi-messenger emitters. Firmamento provides access to a wealth of astronomical data, including imaging, spectral, and timing information, along with tools for data analysis and machine learning. A key feature is the Error Region Counterpart Identifier (ERCI), which helps locate potential counterparts to $\gamma$-ray and other high-energy sources pinpointing objects exhibiting blazar-like characteristics.
Firmamento grants access to spectral data across the entire electromagnetic spectrum, offering well-populated SED generation and VHE detectability estimation. 
The platform also supports user-provided data, allowing researchers to incorporate their own source lists.
}
\maketitle{}

\section{Introduction}
We live in a time where the widespread adoption of open data policies,  adherence to FAIR principles\footnote{https://www.go-fair.org/fair-principles}, and the growing availability of tools providing access to science-ready astronomical data are playing an increasingly important role in research and citizen science. 
As a consequence open data access is widely recognized as a driver of innovation and productivity.
Notable initiatives in  this area include the European Science Cloud (EOSC)\footnote{https://www.eosc-portal.eu}, the Research Data Alliance (RDA)\footnote{https://www.rd-alliance.org},  the Research Infrastructure Cluster (ASTERICS)\footnote{https://www.asterics2020.eu/},  and the Open Universe initiative \citep{OpenUniverse_ESPI} at the United Nations.
\begin{figure*}[t!]
\resizebox{\hsize}{!}{\includegraphics[clip=true]{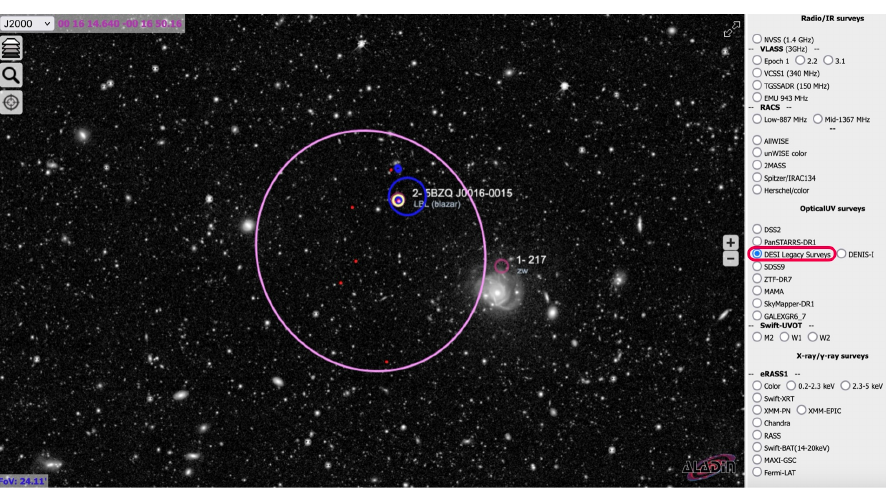}}
\caption{\footnotesize
Example of the Firmamento imaging interface based on Aladin showing the error ellipse of a Fermi-LAT source and the corresponding blazar counterpart superimposed to the DESI Legacy Survey map. Different radio, infrared, optical, UV, X-ray and $\gamma$-ray surveys can be chosen on the right side of the image.}
\label{fig:imaging}
\end{figure*}
In this contribution, we describe Firmamento\footnote{https://firmamento.nyuad.nyu.edu}, a new  online platform dedicated to multi-wavelength astronomical research. It builds on the experience gained through the Open Universe initiative and fully adheres to its principles. 
As such, it aims to assist both researchers and citizens interested in astronomy, making high-energy astrophysics more accessible to a broader audience.

\section{The Firmamento platform}

\begin{figure*}[t!]
\resizebox{\hsize}{!}{\includegraphics[clip=true]{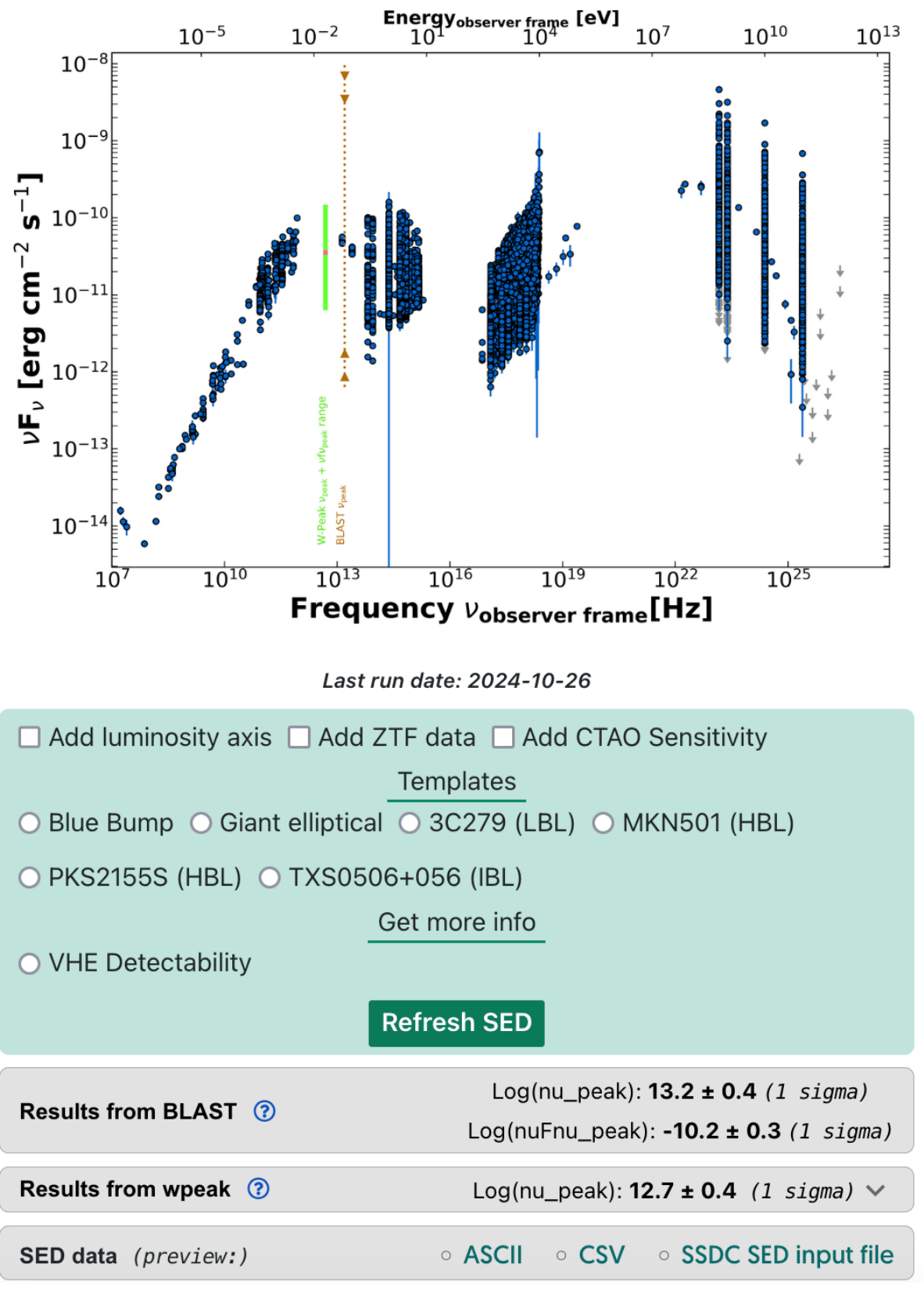}}
\caption{\footnotesize 
An example of SED that is well-populated at most frequencies. The brown and green segments mark the peak of the synchrotron component ($\nu_s$) estimated with the BLAST and Wpeak methods. The $\nu_s$ values are shown in the bottom part of the image.}
\label{fig:SED}
\end{figure*}

Firmamento is a new generation online tool, developed at the New York University-Abu Dhabi, that provides access to astronomical imaging, spectral, timing and tabular  data. Additionally, it incorporates machine learning and other methods to help analyze the data retrieved from about 100 remote and local catalogs and spectral databases.
A full description of Firmamento is provided in \cite{firmamento}. 
In the following, we present a brief overview of the platform, focusing on  aspects that are particularly useful for high-energy astrophysics and distinguish it from other on-line services. These include a) the identification of blazars in $\gamma$-ray and X-ray error regions in  
current and planned observatories such as Fermi-LAT, Swift, eROSITA, CTA, ASTRI Mini-Array, LHAASO, IceCube, KM3Net, SWGO, b) 
the generation of Spectral Energy Distributions (SEDs) and their characterization through the calculation of the synchrotron peak energy and intensity using algorithms and machine learning methods, and c) the prediction of detectability at Very High Energy (VHE) by the current and future observatories. 
The data that Firmamento retrieves  can be used to study the spectral and variability properties of cosmic sources as well as to constrain physical models.
Finally, we briefly describe the results of some recent searches for blazars using Firmamento with data from the Fermi-LAT 4FGL-DR4 catalog of $\gamma$-ray sources and the eRASS1 X-ray sky survey. 
Using the Firmamento ERCI tool, more than 400 previously unassociated Fermi-LAT sources have been identified with plausible blazar counterparts. A similar number of blazars have also been selected among eRASS1 X-ray sources, forming a well-defined statistical sample. This sample will be instrumental in addressing—and likely solving—long-standing questions in high-energy astrophysics that have remained open since the 1990s.

\clearpage
\begin{figure}[t!]
\resizebox{\hsize}{!}{\includegraphics[clip=true]{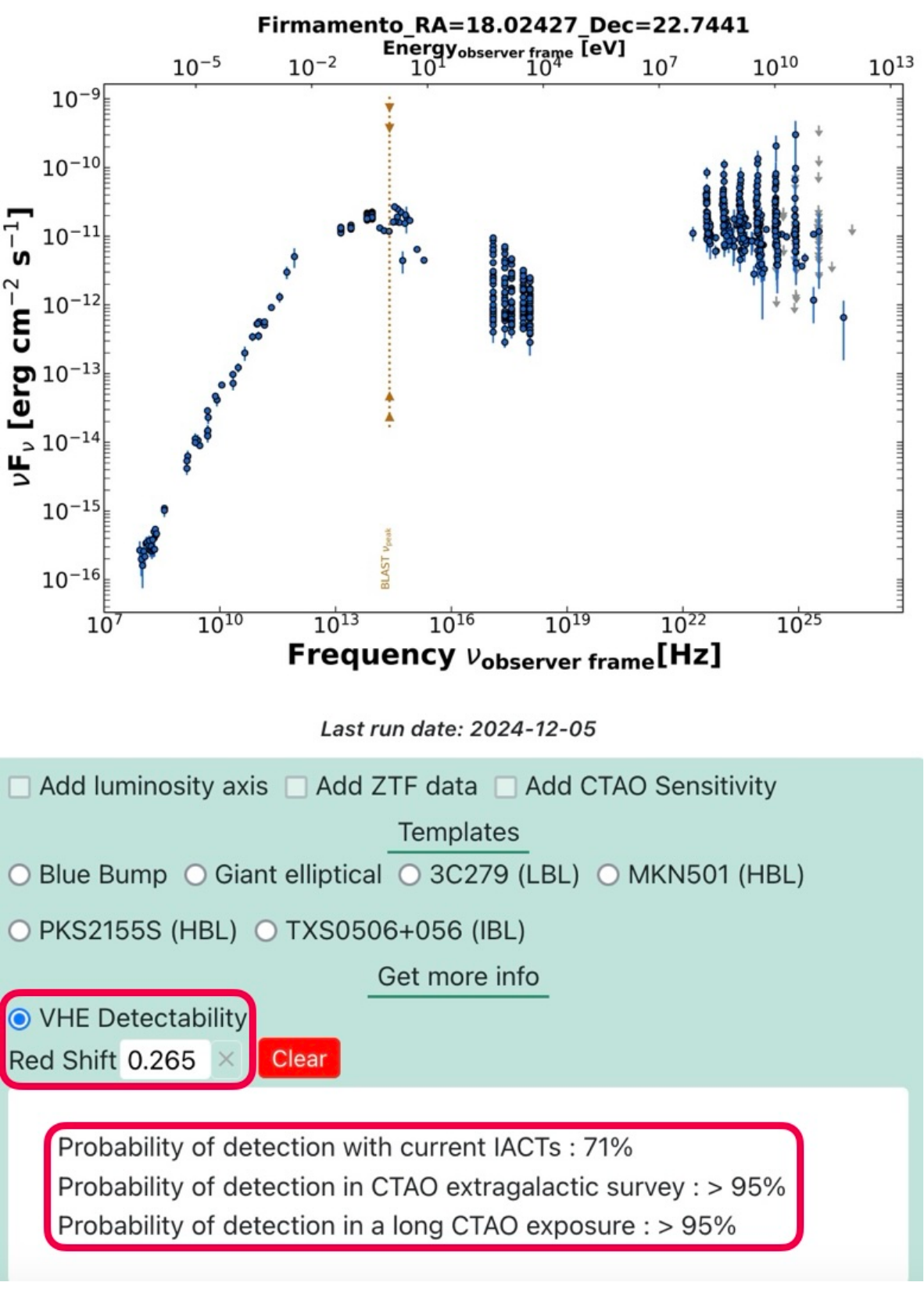}}
\caption{\footnotesize
Example of the VHE detectability estimation tool applied to the blazar 5BZB J0112+2244, located at a redshift of z=0.265z=0.265. The spectral energy distribution (SED) of the blazar is displayed in the top panel of the figure.
}
\label{fig:VHE_detectability}
\end{figure}
\section{The Error Region Counterpart Identifier}

\begin{figure}[t!]
\resizebox{\hsize}{!}{\includegraphics[clip=true]{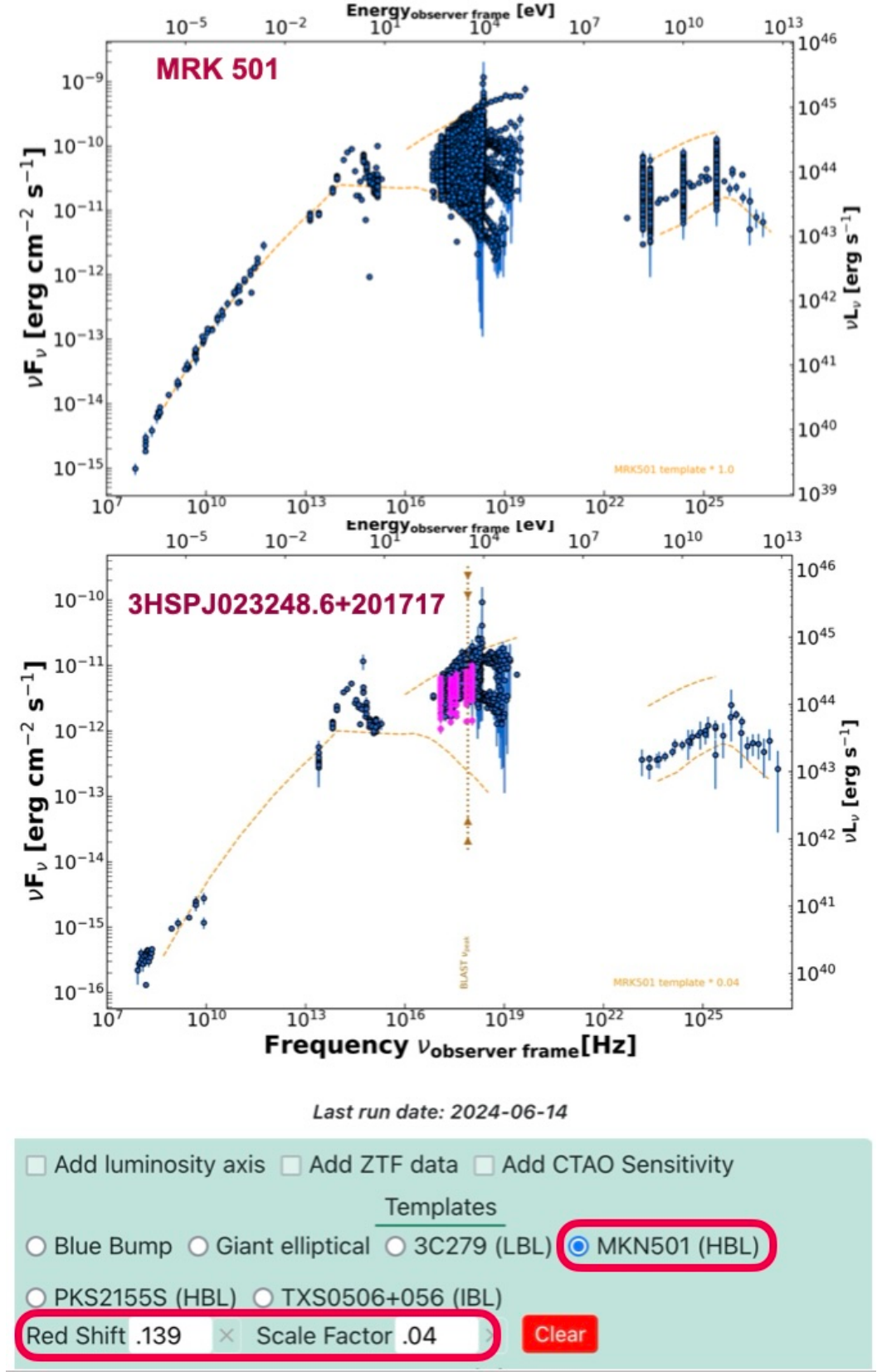}}
\caption{\footnotesize
The MKN501 template is applied to the blazar 
3HSP J023248.6+201717, with a scaling factor of 0.04 (middle image) and, for explanation purposes, to itself with a scaling factor of 1.0 (top SED). The red rectangular areas highlight how to select a template and how to input the required parameters.}
\label{fig:template}
\end{figure}

\begin{figure*}[t!]
\resizebox{\hsize}{!}{\includegraphics[clip=true]{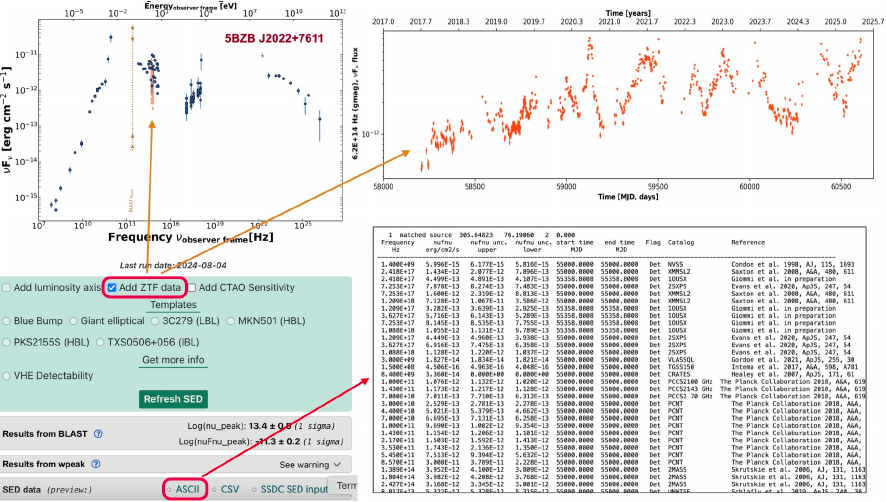}}
\caption{\footnotesize
Optical time-domain data from the Zwicky Transient Facility (ZTF) can be retrieved and plotted in an SED and as a lightcurve by clicking on the "add ZTF data" option. All SED data can also be downloaded in various formats as shown in the bottom right part of the image.}
\label{fig:timing}
\end{figure*}

\begin{figure}[t!]
\resizebox{\hsize}{!}{\includegraphics[clip=true]{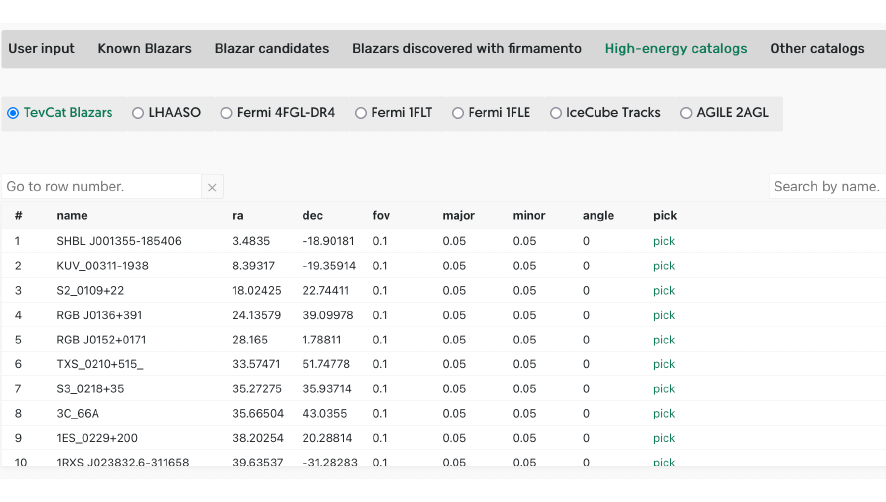}}
\caption{
\footnotesize Firmamento provides tables of sources of different types. This figure shows a table of TeV detected blazars extracted from the TeVCat catalog.
}
\label{fig:catalogs}
\end{figure}

\begin{figure}[t!]
\resizebox{\hsize}{!}{\includegraphics[clip=true]{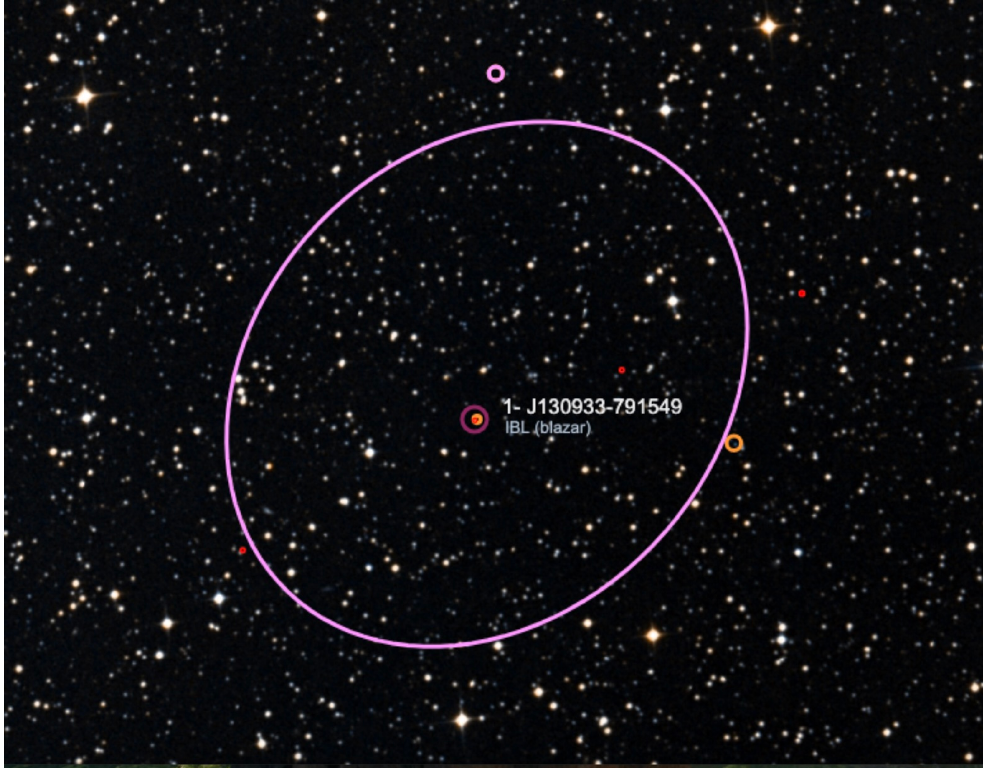}}
\caption{\footnotesize 
Example of the identification of a blazar in a previously unassociated $\gamma$-ray source using ERCI: the case of 4FGLJ1309.4-7915}
\label{fig:4fglJ}
\end{figure}

One of the most important and unique features available within Firmamento is the \textit{Error Region Counterpart Identifier} (ERCI), a tool designed to pinpoint plausible   counterparts in the localization regions of X-ray, $\gamma$-ray, and other high-energy astrophysical sources. This tool is based on an enhanced version of VOU-Blazars \citep{VOU-blazars} specifically developed for Firmamento.
For each source located in an area slightly larger than the error region considered (to account for possible counterparts slightly outside the 95\% localization area), Firmamento retrieves multi-frequency data. 
This includes information on the source’s spatial extension and temporal variability, obtained from approximately 50 remote catalogs.  
For every radio and X-ray source, Firmamento constructs a SED using the available multi-frequency data. 
A dedicated algorithm searches for blazar-like sources by analyzing the SED gradients in key regions, including the radio-to-microwave, infrared-to-optical, optical-to-UV, optical-to-X-ray, and $\gamma$-ray ranges. This method allows the identification of potential blazar candidates by focusing on the shape of their multi-frequency emission patterns and variability.
Particular care is used to identify possible non-jet-like (thermal) components, such as host galaxy or radiation from an accretion disk, which exhibit distinctive signatures in the infrared and optical-UV bands.
The information from the gradients in different parts of the SED, combined with the possible detection of non-jet components, source extension at optical and X-ray energies, and the presence of temporal variability, is used by the ERCI to determine whether a source may be a blazar, another type of AGN, or a Galactic source. 

Fig \ref{fig:imaging} illustrates an example of a $\gamma$-ray error region (purple ellipse) plotted onto a map of the sky from the PanSTARRS survey  along with the ERCI-suggested counterpart, the blazar 5BZQ J0016-0015. The yellow circle represent the Fermi 4LAC-DR3 counterpart, while the blue circle is the error circle of a X-ray source and the red (smaller) circles the position of a radio source.

\section{Imaging Survey Data Across Most Energy Bands}
Firmamento provides access to a wide range of recent and classical imaging sky surveys across the radio, IR, optical, UV, X-ray and $\gamma$-ray energy bands. 
This capability is enabled by the the powerful Aladin\footnote{\url{https://aladin.u-strasbg.fr}}  
viewer \citep{aladin}, which is integrated into Firmamento, as illustrated in Fig. \ref{fig:imaging}.
The figure shows an example of the imaging interface, where the error ellipse of a Fermi-LAT source and its corresponding blazar counterpart are superposed on an optical map from the PanSTARRS-DR1 survey. 
On the right side of the image, there is a menu that allows Firmamento users to select from more than 30 imaging surveys covering radio, infrared, optical, UV, X-ray and $\gamma$-ray frequencies.

\section{The SED generator and peak estimators}

Firmamento retrieves multi-frequency data across the entire electromagnetic spectrum via an enhanced version of
VOU-Blazars \citep{VOU-blazars}, which incorporates
more catalogs than those included in earlier releases.

Once the output files from each catalog providing data for the specified position are retrieved, a module converts the different formats and local units into monochromatic
$\nu$ F$_\nu$ values. It then applies the appropriate de-reddening corrections at near infrared, optical and X-ray frequencies, to take into account absorption within the Galaxy.
The corrected $\nu$F$_\nu$ values are then combined to construct a full SED, as illustrated in Fig. \ref{fig:SED}, for the well-known and frequently observed bright blazar 3C454.3.
This SED building method is being developed in collaboration with the Markarian Multiwavelength Data Center \citep[MMDC,][]{MMDC}, where the output SED data can be  fit to theoretical blazar models. 

\subsection{The BLAST and W-peak synchrotron peak estimators}

Once the SED of a blazar or candidate blazar is available, Firmamento estimates the synchrotron peak frequency ($\nu_s$), and the flux
at peak ($\nu_s F (\nu_s)$), using the BLAST and the W-peak tools.
BLAST \citep{Blast} is a machine-learning-based
estimator designed for the automated
estimation of the peak frequency and peak intensity directly from the observed SED data points. W-peak is a
tool that estimates the synchrotron peak frequency and
peak flux by analyzing the infrared spectral slopes from
the WISE and NEOWISE data. These are particularly predictive of blazar behavior if the data can be
attributed solely to the jet, without contamination from
by the host galaxy, the accretion disk, dusty torus, or
the broad line region \citep{wpeak,firmamento}.
\subsection{The VHE detectability estimator}
Once a SED is available with a reasonable amount of multi-frequency data to reliably calculate $\nu_s$ and $\nu_sF_{\nu}$, and the redshift of the blazar is known, Firmamento can estimate the probability of detection at VHE $\gamma$-ray energies,  based on the method described in \citet{wpeak}.
By applying this technique, Firmamento enables researchers to predict the detectability of sources at VHE energies, facilitating targeted follow-up observations by current and future VHE $\gamma$-ray observatories.

Fig. \ref{fig:VHE_detectability} illustrates the case of the application of this method to the blazar 5BZB J0112+2244, located at a redshift z = 0.265. According to the VHE detectability tool this object has a probability of detection  of 71\% with the current generation of Imaging Atmospheric Cherenkov Telescopes (IACTs).
Furthermore, the probability of detection exceeds 95\% with the upcoming Cherenkov Telescope Array Observatory (CTAO\footnote{\url{https://www.ctao.org/}}).

\subsection{Blazar SED templates}

It is often useful to compare the SED of an object with those of well-known blazars of different types. This can be done in Firmamento using the overlay template option, as illustrated in Fig. \ref{fig:template}. 
Four types of well-known blazar templates are available, representing different SED synchrotron peak energy values: 3C279 for a low energy $\nu_s$, TXS0506+056 for intermediate values of $\nu_s$, PKS2155-304 for high values of $\nu_s$, and MKN501 for very high values of $\nu_s$.
For clarity and to illustrate how the templates (orange dotted lines) are constructed, the top part of Fig. \ref{fig:template} shows the MKN501 template applied to its own SED with a scaling factor of 1.0. The same template, rescaled by a factor of 0.04 is plotted on the SED of the blazar 
3HSP J023248.6+201017 in the middle part of the figure.

\section{Time-domain data}

One data domain often overlooked in astronomical web services is time.
Firmamento, alongside intensity, provides the observation date for each data point, whenever it is available in the remote archives. 
This type of information can be used to build light curves at all 
frequencies, provided that enough data points are available.
Firmamento also provides transparent interfaces to sites where timing information is offered. One such example is the interface with the Zwicky Transient Facility (ZTF), which is shown in Fig. 
\ref{fig:timing} where the lightcurve of the blazar 5BZB J2022+7611 based on the ZTF optical data in the g filter, converted to $\nu F\nu$ units, is shown on the right.  

\section{Firmamento catalogs and users tables}
Firmamento provides access to tables and catalogs of known astronomical objects of a variety of types, including a list of over 6,400 blazars, and several lists of high-energy detected sources. As an example Fig. \ref{fig:catalogs}
shows a table of TeV
detected blazars extracted from the TeVCat catalog\footnote{\url{https://www.tevcat.org/}} and the possibility to choose other lists of high-energy detected sources.  
In addition, Firmamento users can provide their own list of sources by uploading a simple comma-separated-values (csv) file. The file must include the source name, R.A. and Dec. (in degrees and J2000), along with optional parameters such as the search radius around the source and the  uncertainty region, which is assumed to have an elliptical shape. A comment field can also be added. The following is an example of the required format:

{\scriptsize
\begin{verbatim}  
Name,ra,dec,fov,major,minor,angle,comment
source1,12.516,1.579,0.3,0.2,0.1,4.,comment1
source2,13.488,1.544,0.4,0.2,0.2,0.,comment2
...
\end{verbatim}
}
The parameter "fov, major, minor, angle" are the size (in arcminutes) of the search area including the uncertainty region and the parameters of the ellipse (in arcminutes for the axes and degrees for the rotation angle). If the file only includes the fields "name, ra and dec;" that is, if "fov," "major," "minor" and "angle" are missing, then Firmamento assumes that the position of the source has no uncertainty and it will only generate the SED using the multifrequency data that match the specified position.

\section{Some Firmamento-based scientific results}

\subsection{The FLAC $\gamma$-ray catalog}
The Firmamento ERCI tool was used by a team of researchers and students to conduct an independent study on high-Galactic-latitude sources from the Fermi-LAT 4FGL-DR4 $\gamma$-ray sources catalog (Giommi et al. 2025, submitted).
The results agree
with those of the Fermi 4FGL-DR4 and 4LAC-DR3 catalogs in the large majority of the cases, but differ in a significant fraction. In particular, the authors have been able to find credible blazar counterparts for over 400 previously unassociated $\gamma$-ray sources, reducing the fraction of still unassociated extragalactic Fermi-LAT sources, to less than 20\%.
Fig. \ref{fig:4fglJ} shows an example of ERCI identification  of a blazar in the error ellipse of the $\gamma$-ray source 4FGLJ1309.4-7915, a previously unassociated $\gamma$-ray source. 
For each blazar associated to a 4FGL $\gamma$-ray source Firmamento
calculates its $\nu_s$ and $\nu_sF_{\nu_s}$.
These crucial SED parameters are used to a) subclassify the blazar into one of three now standard categories: low-energy peaked (LBL), intermediate-energy peaked (IBL), or high-energy peaked (HBL) blazar, and b) estimate the source total flux or luminosity.  
This work resulted in the compilation of a new $\gamma$-ray source counterparts catalog
called 1FLAC, which is available through Firmamento.

\subsection{The BreRASS X-ray survey}
The BlazaR eRASS (BReRASS, Giommi et al. 2025, in preparation) survey consists of a  well-defined, carefully selected 
sample of blazars discovered among eRASS1 \citep{erass1} and eFEDS \citep{efeds} X-ray sources using Firmamento.  To efficiently find blazars the authors  cross-matched the position of eRASS1 and eFEDS sources with the NVSS, VLASS and RACS radio surveys.
The Firmamento ERCI was then used to identify the different types of X-ray sources in the preliminary sample and select blazars with an efficiency $>$ 80\%.
The resulting list of blazars, was carefully inspected by the authors, leading to the selection of over 500 confirmed objects, a sample that
is significantly larger than those in previous studies. This  new sample will enable the authors to conduct a highly detailed analysis, with the potential to address long-standing questions—such as the precise X-ray LogN-LogS relation and the cosmological evolution of BL Lacs—that have remained open since the 1990s.

\section{Conclusions}
We briefly described the characteristics of Firmamento, a new web-based  tool dedicated to blazars and multifrequency/multimessenger emitters, focussing on features most relevant to high-energy astrophysics.
Today, many aspects of modern astrophysics rely heavily on multifrequency data. This new facility aims to become a valuable service in supporting $\gamma$-ray astronomy, providing tools and resources to enhance the identification and characterization of high-energy sources.
It is also notable that Firmamento employs algorithms and machine-learning tools to assist professional scientists in innovative ways. Additionally, it aims to make astrophysical research more accessible to a broader audience, including individuals with varying levels of expertise.
The results of ongoing research projects, such as the FLAC catalog and the BreRASS survey, demonstrate that Firmamento is actively contributing to high-energy astrophysics. It helps to characterize thousands of blazars and discover hundreds of new ones, while also generating well-defined samples of sources suitable for unbiased statistical studies.



\begin{acknowledgements}
The author expresses his gratitude to the Center for Astrophysics and Space Science (CASS) of New York University, Abu Dhabi, for supporting the development of Firmamento  and his research visits at NYU-Abu Dhabi.

\end{acknowledgements}


\begin{thebibliography}{51}
\bibitem[Bonnarel et al.(2000)]{aladin} Bonnarel, F., Fernique, P., Bienaym{\'e}, O., et al.\ 2000, \aaps, 143, 33. doi:10.1051/aas:2000331
\bibitem[Chang et al.(2020)]{VOU-blazars} Chang, Y.-L., Brandt, C.~H., \& Giommi, P.\ 2020, Astronomy and Computing, 30, 100350. doi:10.1016/j.ascom.2019.100350
\bibitem[Giommi et al.(2024)]{wpeak} Giommi, P., Sahakyan, N., Israyelyan, D., et al.\ 2024, \apj, 963, 48. doi:10.3847/1538-4357/ad20cb
\bibitem[Giommi et al.(2018)]{OpenUniverse_ESPI} Giommi, P., Arrigo, G., Barres De Almeida, U., et al.\ 2018, arXiv:1805.08505. doi:10.48550
\bibitem[Glauch et al.(2022)]{Blast} Glauch, T., Kerscher, T., \& Giommi, P.\ 2022, Astronomy and Computing, 41, 100646. doi:10.1016/j.ascom.2022.100646
\bibitem[Merloni et al.(2024)]{erass1} Merloni, A., Lamer, G., Liu, T., et al.\ 2024, \aap, 682, A34. doi:10.1051/0004-6361/202347165
\bibitem[Sahakyan et al.(2024)]{MMDC} Sahakyan, N., Vardanyan, V., Giommi, P., et al.\ 2024, \aj, 168, 289. doi:10.3847/1538-3881/ad8231
\bibitem[Salvato et al.(2022)]{efeds} Salvato, M., Wolf, J., Dwelly, T., et al.\ 2022, \aap, 661, A3. doi:10.1051/0004-6361/202141631
\bibitem[Tripathi et al.(2024)]{firmamento} Tripathi, D., Giommi, P., Di Giovanni, A., et al.\ 2024, \aj, 167, 116. doi:10.3847/1538-3881/ad216a

\end{thebibliography}
\end{document}